\documentclass[mathleft
]{an}
\usepackage{graphicx}
\usepackage{times}
\usepackage{ulem}
\begin{document}

\Pagespan{789}{}
\Yearpublication{2006}%
\Yearsubmission{2005}%
\Month{11}%
\Volume{999}%
\Issue{88}%

\title{The outburst properties of AM CVn stars}

\author{Iwona Kotko\inst{1}\fnmsep\thanks{Corresponding author:
  \email{ikotko@nac.oa.uj.edu.pl}\newline}
\and Jean-Pierre Lasota\inst{1,2}
\and Guillaume Dubus\inst{3}
}
\titlerunning{Outbursts of AM CVn}
\authorrunning{Kotko et al.}
\institute{
Astronomical Observatory, Jagiellonian University, Cracow, Poland
\and
Institut d'Astrophysique de Paris, UMR 7095 CNRS, UPMC Univ Paris 06, Paris, France
\and
Laboratoire d'Astrophysique de Grenoble, INSU/CNRS, UMR 5571, Universite Joseph Fourier, Grenoble, France}

\received{30 May 2005}
\accepted{11 Nov 2005}
\publonline{later}


\abstract{
  We briefly summarize the observational properties of ultra-compact binaries called AM CVn stars. We analize their outbursts originating from the thermal-viscous instability in helium accretion disc. We present our preliminary results in applying the model of Dwarf Novae outbursts to helium discs. We can calculate models of outbursts of reasonable amplitude of 2 mag with a constant $\alpha$ parameter throughout the calculation. Setting the mass transfer rate close to its upper critical value produces model lightcurves that resemble short superoutbursts.  }

\maketitle

\section{Introduction}
AM CVn stars are very close binary systems, consisting of two white dwarfs. Because of their proximity mass is transfered from a donor star and, unless the combined effect of small orbital separation and size of the accreting white dwarf prevent it (see Nelemans 2005), an accretion disc forms. Observations show that spectra of these binaries are completely deprived of hydrogen lines and that helium is the dominant chemical component both in the donor's and disc's composition. Another important, observational feature characterizing AM CVn stars is their ultra-short orbital period spanning a range of 10 to 65 minutes.

AM CVn stars can be observed in three luminosity states: bright-persistent, faint-persistent and outbursting. Among about 20 objects identified as AM CVn, 5 systems were observed to be outbursting.

\section{A few words about Dwarf Novae}
\subsection{Theoretical approach}

Dwarf Novae are hydrogen-dominated Cataclysmic Variable stars (CVs; see Warner 1995) that exhibit outbursts with typical amplitudes of 2-5 mag and 2-20 days duration. In addition to such {\sl normal} outbursts, some systems show so-called {\sl superoutbursts} of larger amplitudes and longer duration (some CVs have only rare superoutbursts).

The Disc Instability Model (DIM; see Lasota 2001 for a review) attempts to describe the mechanism of disc outbursts in close binaries. The general idea is that they are caused by a thermal-viscous instability in the disc which is triggered by changes in the opacities, occurring at temperature of ionization of the disc's dominant chemical element, while the material in the disc becomes partially ionized.

The thermal-equilibrium solutions of the set of differential equations describing the vertical structure of a particular ring of a disc form an S-shaped curve on the effective temperature -- surface density plane. The middle branch of this curve, with a negative slope, represents thermally unstable states. The lower branch is where disc is cold and stable (cold branch) and the upper one is where the disc is hot and stable (hot branch). 

The first simulations of disc outbursts showed that when one assumed throughout the S-curve a constant value of the Shakura-Sunyaev (1973) viscosity $\alpha$-parameter one got only low-amplitude oscillations in the lightcurve, instead of outbursts of observable amplitude. It was soon realized by Smak (1984) that in order to produce observed amplitudes with the DIM one has to take different values of $\alpha$ for the cold and hot branches of the S-curve. In this way one extends the range between the critical surface densities defining the middle part of the S-curve allowing the outbursts to reach higher amplitudes. One should stress, however, that this ansatz is purely phenomenological and the physical mechanism for such a change is unknown.
Tsugawa and Osaki (1997) showed that changing the metallicity of the disc has a similar effect of middle-branch stretching: a purely helium disc with constant $\alpha$ will have a longer middle branch than a hydrogen disc for the same set of parameters.
\subsection{Observational features}

Just after the first observations of the lightcurves of outbursting AM CVn stars it has been realized that these systems may be the helium cousins of hydrogen-dominated Dwarf Novae.
Dwarf Novae are classified into three groups, according to the character of their outbursts:

\begin{itemize}
\item[a).]
 \textbf{SU UMa -type}\\
In the lightcurves of this group, apart from normal outbursts, one observes so-called superoutbursts. They are a few times longer than the normal ones and brighter than them by about 0.7 magnitude. Superoutbursts are accompanied by superhumps, i.e. low-amplitude periodic modulations in the light emitted from the binary. The superhump period differs by few percent from the orbital period.

Among SU UMa's one finds two special types of systems:
  \begin{itemize}
  \item[--]
\textbf{WZ Sge -type}: These Dwarf Novae show superoutbursts only with very long recurrence times between them (at least several years). During the decay from maximum the luminosity of these systems rises again instead of decreasing to the quiescence level and enters a phase during which it varies for about 1 magnitude with about 2 day repetition period. After a few such cycles the system light falls to minimum state (Patterson et al. 2002b). 
  \item[--]
 \textbf{ER UMa-type}: which have very short sypercycles, below 50 days.
  \end{itemize}
\item[b).]
\textbf{Z Cam} 

Their characteristic feature is standstills, which are the periods after outbursts lasting for several days with 0.5 mag fainter luminosity than during outbursts. After that systems return to the quiescent, faint state.
\item[c).]
\textbf{U Gem} 

These systems are supposed to show normal outbursts only, but the prototype itself was observed to have one superoutburst (Smak \& Waagen 2004) in more than 100 years of recorded observation.

\end{itemize}

\section{Properties of AM CVn outbursts}
In contrast to these detailed records, among the 5 outbursting AM CVn stars only the lightcurves of CR~ Boo and V803~Cen are thorough enough to allow drawing conclusions about the character of the outbursts, which are very complex.

\subsection{CR Boo}

In the lightcurve of this AM CVn binary one clearly distinguishes superoutbursts. The fact that additionally confirms the nature of superoutbursts is the presence of superhumps in the light of CR Boo at its brightness maximum. That allows to claim that it is helium version of SU UMa type Dwarf Nova. The length of its supercycle was estimated to be 46.3 days in 2000 (Kato et al. 2000). But one year later it surprisingly appeared that it had changed and from a set of observations on June 2001 Kato et al. estimated that it lasted only 14.7 days. This makes reasonable to associate CR~Boo with RZ~LMi and DI~UMa - the binaries of ER~UMa type with extremely short sypercycles in the range of 19-25 days.

The object becomes fainter after the superoutburst, however, it does not return to its quiescence luminosity brightening instead again and entering a so called cycling phase. This phase is very similar to so called ``echo outbursts" seen in only one type of hydrogen Dwarf Novae -- the WZ Sge. The oscillations have an amplitude of about 2 mag and are about 1 mag fainter at their maximum than the superoutburst. At minimum they are 2.5-3 mag brighter than in quiescence. After several cycles the system returns to quiescence.

\subsection{V803 Cen}

Superoutbursts with superhumps are also characteristic features of V803 Cen outbursts, but they are longer than in the case of CR Boo. Kato et al. (2004) described their properties in detail. Superoutbursts of V803 Cen last for 3-12 days, their brightness linger around V=12.7-13.3 and the whole supercycle lasts for about 77-85 days. Again there is a reason to think about V803 Cen as a helium SU~UMa type object. The situation becomes more complicated due to what happens to its lightcurve after the superoutburst. During observations in April 1997 and June 2000 the star instead of fading back to its faint state ($\approx$ V=17) after superoutburst, remained on the level of 13.3-14.0 mag and stayed there for a longer period of time, which is unfortunately difficult to determine due to an observational gap. This period was identified as a state similar to standstill characteristic for Z~Cam stars. However, here one observes 1 magnitude modulations which are absent in the classical standstills in Z Cam group.
In June 2003 a cycling state similar to that of CR~Boo was identified. Looking at this part of the V803~Cen lightcurve one notices striking similarity to the lightcurve of WZ~Sge, with the difference that in the case of V803~Cen the phase of superoutburst is much shorter than for WZ~Sge and the amplitude of ``reflare" outbursts is much lower: it is about 1.4 magnitude. 

This behaviour is also characteristic of other observed superoutbursts of this object.

\begin{figure*}
\centering
\includegraphics[width=0.65\textwidth, angle=-90]{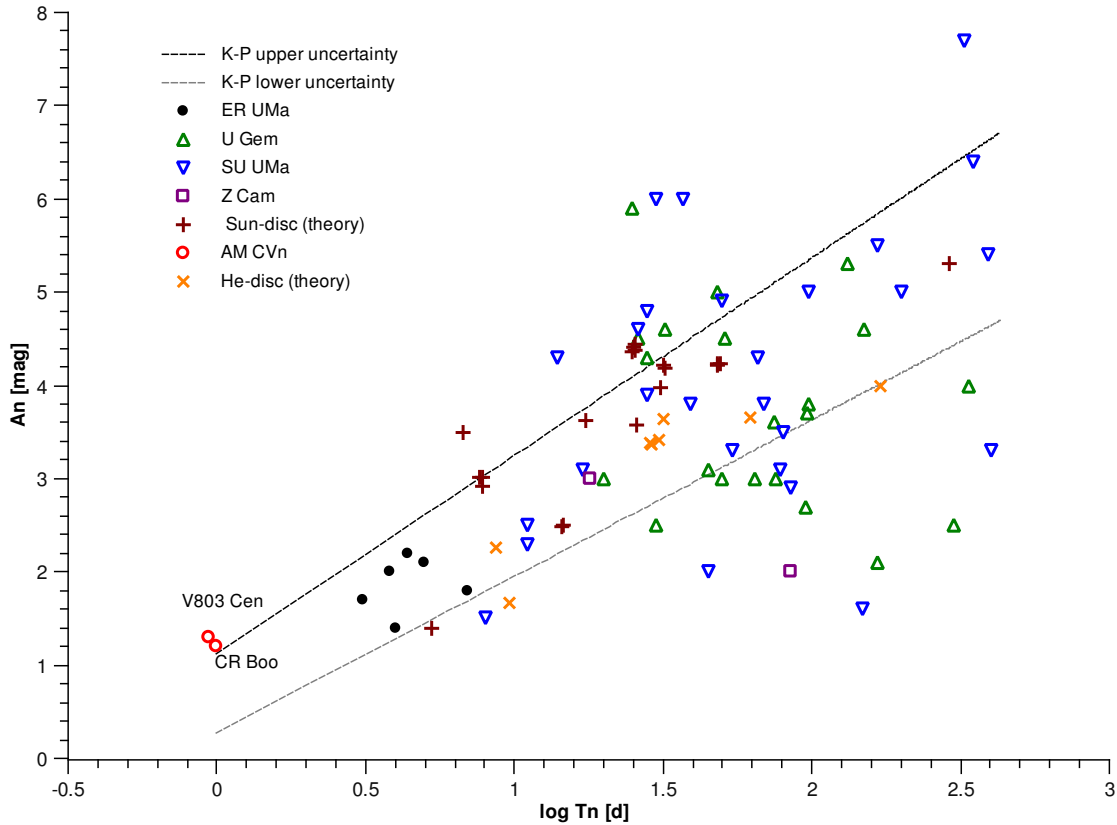}
\caption{Kukarkin-Parenago relation for real objects and model lightcurves.}
\label{K-P realtion}
\end{figure*}

\section{The Kukarkin-Parenago relation}

Patterson et al. (2000) identified the cycling part of the lightcurves of CR Boo and V803 Cen with normal outburst of Dwarf Novae. They based this statement on the Kukarkin-Parenago relation, that connects the amplitude of an outburst with the duration of the outburst cycle which is sometimes called the outburst ``period" (i.e. recurrence time). This relation was derived by Kukarkin and Parenago (1934, hereafter K.-P.) for Dwarf Novae on basis of their Variable Stars catalogue. (See Warner 1995 for an updated version.) This is a completely empirical relation, resulting from a linear fit to points marked on the amplitude-period diagram.

Patterson et al. (2000) found that amplitudes and period of the ``rebrightenings" in CR Boo and V803 cycling states match well the above mentioned relation.

In Figure 1 we present the amplitudes and periods of normal outbursts for several types of systems and the K.~-~P. relation. We marked there all SU UMa (down triangles) and U Gem (up triangles) type Dwarf Novae from the catalogue by Ritter and Kolb (2003; 2009) with known amplitudes and periods of their normal outbursts. There are also two squares representing Z Cam type binaries. These were the only ones with determined amplitude and outburst period. The black points are all observed ER UMa class binaries. As can be seen they lie within the lines which correspond to relation derived originally by Kukarkin and Parenago and updated by Warner (1995) :
$$A_{\rm n}=0.7 \pm 0.43 + (1.9 \pm 0.22)\rm logT_{\rm n}$$ 
The relation has estimated uncertainties for coefficients and so taking them into account resulted in 2 lines on the diagram instead of one liear fit. 

To test if outbursts produced by model simulations would fulfill the K.-P. relation we created synthetic lightcurves for different sets of parameters for solar composition discs and helium discs.
We took into account 2 values of primary mass : $0.7M_{\odot}$ and $1.4M_{\odot}$, 2 values of $\alpha_{\rm hot}$ with 4 values of $\alpha_{\rm cold}$ and 3 values of disc outer radius $R_{\rm disc}: 2\times10^{10}, 3\times10^{10}, 5\times10^{10}$ cm in various configurations. The parameters kept constant throughout calculations were the inner disc radius $R_{\rm in}=8.5\times10^8$ and mass transfer rate from secondary $\dot M=10^{16}$ g/s. The results are presented as crosses. As can be seen points from simulations lie close to the empirical lines. This confirms that the DIM is indeed describing outbursts of Dwarf Novae.

\section{Results}

Motivated by the statement of Patterson et al. (2000), that outbursts in ``cycling" states in V803 Cen and CR Boo lightcurves are normal outbursts as in Dwarf Novae, we decided to concentrate on the cycling part of lightcurves at first. Our aim was to reproduce the alleged normal outbursts seen in the lightcurves of V803 Cen and CR Boo with DIM applied to helium disc.

The standard approach with $\alpha_{\rm cold}<\alpha_{\rm hot}$ failed since instead of outbursts with amplitude of about 2 mag, we obtained amplitudes of 5-6 mag. Keeping in mind the fact that for pure helium discs the unstable part of the S-curve covers a wider range in surface densities than for hydrogen discs we decided to return to the most basic version of the model and to keep the viscosity parameter constant for all states of the disc. While in the hydrogen case applying $\alpha_{\rm cold}=\alpha_{\rm hot}=\alpha_{\rm const}=0.1$ we obtained only about 0.5~mag oscillations, for helium discs the outburst amplitudes are similar to the observational values.

The next step was to run the code (Hameury et al. 1998) with different values of mass transfer rate from secondary star. The other parameters were kept unchanged. In this way we managed to reproduce also the observed shape of the cycling part of investigated lightcurves. The thing which still remained unsatisfying was the period of outbursts being about 10 times too long. So in order to shorten it we tried to change $\alpha_{\rm const}$ from 0.1 to 0.5. Comparing our final result (upper graph in Fig.2 ) with the observational lightcurve of V803 Cen kindly provided by Patterson (middle graph in Fig.2) one sees similarity in shape, amplitude and outburst period between them. Our further calculations should lead to better adjusting the height of amplitude to observational 1.5-2 mag. Probably the point is to change the metallicity in the model disc by passing from purely helium disc to disc mainly composed of helium with admixture of heavier elements such as nitrogen. Higher metallicity should lower the outbursts amplitude. It would be also closer to real situation. It is known from spectroscopic observations of AM~CVn stars that in addition to helium they show also N, Si or Ne lines (Nelemans 2005).

The interesting point is that for values of mass transfer rate close to the critical value $\dot {M^+}_{\rm crit}$ above which disc in our simulations becomes hot and stable, the lightcurves show a plateau and change their shape from narrow to wide. They start to resemble short superoutbursts. It is worth to point out that it happens without increasing mass transfer rate during the outburst, i.e. without secondary irradiation. This doesn't happen in the simulations for the hydrogen dominated disc.

\begin{figure}
\includegraphics[width=83mm,height=85mm]{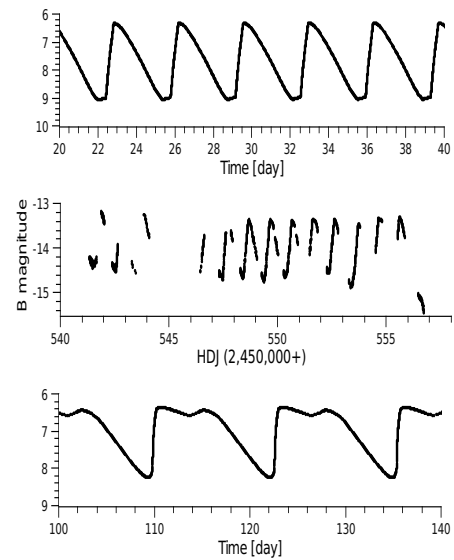}
\caption{{\sl Top}: the lightcurve obtained from the model for $\alpha_{\rm cold}=\alpha_{\rm hot}=0.5$ and mass transfer rate $\dot M=3\times10^{16} \rm g/s$. {\sl Middle}: Observational lightcurve of V803 Cen. {\sl Bottom}: model with $\alpha_{\rm cold}=\alpha_{\rm hot}=0.1$ and mass transfer rate $\dot M=5.2\times10^{17} \rm g/s$ }
\label{Our results}
\end{figure}

\section{Conclusions}

Our preliminary results suggest that the DIM applied to helium discs can reproduce some properties of outbursting AM CVn's. Setting $\alpha$ constant and adjusting the mass transfer rate we are able to reproduce cycling part of the lightcurves of CR Boo and V803 Cen. However, this success is purely formal as the physical reason for such a aboundance dependence of the viscosity parameter is not understood at all. More interesting is the result that sufficiently high mass transfer rate results in wide outbursts, resembling standstills in V803 Cen lightcurve, without the need of increasing mass transfer rate during the rise of an outburst.

Our further attempts will address obtaining other features of observational lightcurves of AM CVn such as superoutbursts or standstills.
It would be very instructive to obtain more measurements of lightcurves from observations in order to understand behaviour and nature of these interesting objects.

\acknowledgements

We thank Joe Patterson for providing us with the CR Boo and V803 Cen data. This work was supported in part by the Polish Ministry of Science and Higher Education project N N203 380336, by the French Space Research Centre CNES and by the European Commission via contract ERC-StG-200911.

\end{document}